\documentclass[sigconf]{acmart}

\usepackage{listings}
\usepackage{color}
\usepackage{booktabs}
\usepackage{ulem}
\usepackage[utf8]{inputenc}
\usepackage[T1]{fontenc}
\usepackage{listingsutf8}
\usepackage{xcolor}
\usepackage{multirow}
\usepackage{tcolorbox}
\usepackage{xspace}

\definecolor{lightgray}{rgb}{.9,.9,.9}
\definecolor{darkgray}{rgb}{.4,.4,.4}
\definecolor{purple}{rgb}{0.65, 0.12, 0.82}

\lstdefinelanguage{JavaScript}{
  keywords={typeof, new, true, false, catch, function, return, null, catch, switch, var, if, in, while, do, else, case, break},
  keywordstyle=\color{blue}\bfseries,
  ndkeywords={class, export, boolean, throw, implements, import, this},
  ndkeywordstyle=\color{blue}\bfseries,
  identifierstyle=\color{black},
  sensitive=false,
  comment=[l]{//},
  morecomment=[s]{/*}{*/},
  commentstyle=\color{purple}\ttfamily,
  stringstyle=\color{red}\ttfamily,
  morestring=[b]',
  morestring=[b]"
}

\colorlet{punct}{red!60!black}
\definecolor{background}{HTML}{EEEEEE}
\definecolor{delim}{RGB}{20,105,176}
\colorlet{numb}{magenta!60!black}

\lstdefinelanguage{json}{
   backgroundcolor=\color{lightgray},
   extendedchars=true,
   basicstyle=\footnotesize\ttfamily,
   showstringspaces=false,
   showspaces=false,
   numbers=left,
   numberstyle=\footnotesize,
   numbersep=5pt,
   tabsize=2,
   breaklines=true,
   showtabs=false,
   captionpos=b,
   inputencoding=utf8,
   literate=
     *{0}{{{\color{numb}0}}}{1}
      {1}{{{\color{numb}1}}}{1}
      {2}{{{\color{numb}2}}}{1}
      {3}{{{\color{numb}3}}}{1}
      {4}{{{\color{numb}4}}}{1}
      {5}{{{\color{numb}5}}}{1}
      {6}{{{\color{numb}6}}}{1}
      {7}{{{\color{numb}7}}}{1}
      {8}{{{\color{numb}8}}}{1}
      {9}{{{\color{numb}9}}}{1}
      {:}{{{\color{punct}{:}}}}{1}
      {,}{{{\color{punct}{,}}}}{1}
      {\{}{{{\color{delim}{\{}}}}{1}
      {\}}{{{\color{delim}{\}}}}}{1}
      {[}{{{\color{delim}{[}}}}{1}
      {]}{{{\color{delim}{]}}}}{1},
}

\AtBeginDocument{%
  \providecommand\BibTeX{{%
    \normalfont B\kern-0.5em{\scshape i\kern-0.25em b}\kern-0.8em\TeX}}}

\setcopyright{acmcopyright}
\copyrightyear{2020}
\acmYear{2020}

\acmConference[ASE 2020]{The 35th IEEE/ACM International Conference on Automated Software Engineering}{21 - 25 September, 2020}{Melbourne, Australia}

\begin{document}

\title{Code-based Vulnerability Detection in \\ Node.js Applications: How far are we?}

\author{Bodin Chinthanet}
\affiliation{%
  \institution{Nara Institute of Science and Technology, Japan}
}
\email{bodin.chinthanet.ay1@is.naist.jp}

\author{Serena Elisa Ponta, Henrik Plate,  Antonino Sabetta}
\affiliation{%
  \institution{SAP Security Research, France}
}
\email{serena.ponta@sap.com}
\email{henrik.plate@sap.com}
\email{antonino.sabetta@sap.com}

\author{Raula Gaikovina Kula, Takashi Ishio, Kenichi Matsumoto }
\affiliation{%
  \institution{Nara Institute of Science and Technology, Japan}
}
\email{raula-k@is.naist.jp}
\email{ishio@is.naist.jp}
\email{matumoto@is.naist.jp}

\renewcommand{\shortauthors}{Chinthanet et al.}

\begin{abstract}
With one of the largest available collection of reusable packages, the JavaScript runtime environment Node.js is one of the most popular programming application.
With recent work showing evidence that known vulnerabilities are prevalent in both open source and industrial software, we propose and implement a viable code-based vulnerability detection tool for Node.js applications.
Our case study lists the challenges encountered while implementing our Node.js vulnerable code detector.
\end{abstract}
\maketitle

\newcommand{\company}{SAP}
\newcommand{\tool}{\textit{Eclipse-Steady}\xspace}
\newcommand{\codeAnalysis}{bill of materials~}

\section{Introduction}
As of 2020, the Node.js package manager (i.e., npm) is reported to serve over 1.3 million packages to roughly 12 million developers, who download such packages 75 billion times a month, and all at a growing rate \cite{Web:npmStat}.
Furthermore, as evidence of its influence, the industry giant Microsoft's GitHub had completed its acquisition of npm earlier in April, 2020.
Recent studies have shown evidence that known vulnerabilities can affect both open source and industrial applications alike \citep{Web:ossra}.

Most detection methods for vulnerabilities has been at meta-detection \citep{Web:npm_audit, Web:github_security_noti}.
Meta-detection capabilities rely on the assumption that the metadata associated to Open Source Software (OSS) libraries (e.g., name, version), and to vulnerability descriptions (e.g., technical details, list of affected components) are always available and accurate. 
The metadata, which are used to map each library onto a list of known vulnerabilities that affect it, are often incomplete, inconsistent, or missing altogether. 
Existing works show that such approaches are unreliable and suffer from false positives \citep{Zapata:ICSME2018}.

\citet{Ponta:icsme2018,Ponta:emse2020} proposed \tool, a code-centric and usage-based approach to detect open source vulnerabilities. The \tool~project \citep{Web:Steady} is able to identify, assess and mitigate open source dependencies with known vulnerabilities for Java and Python industry grade applications. It supports software development organizations in regard to the secure use of open source components during application development. 
A code-centric approach reduces the number of false positives and false negatives as it accounts for the actual presence of vulnerable constructs (i.e., constructs that are modified by the patch), no matter where they occur \cite{Ponta:icsme2018,Ponta:emse2020}. Having identified the vulnerable constructs, it is then possible to establish whether they are reachable in the context of an application thereby assessing the potential impact of the vulnerability.

Although most studies use meta-detection (i.e., checking the package.json configuration file) for mapping npm vulnerabilities to the packages, there is yet to be a code-centric approach designed for JavaScript Node.js applications \citep{Decan:2018, Decan:ICSME:2018, Kikas:2017, Zerouali:ICSR2018}.
\citet{Lauinger2016ThouSN} provides evidence that JavaScript issues are prevalent in most web applications, strengthening the argument for a Node.js code-centric solution.
Due to the dynamic event-based nature of JavaScript code, the performance of a code-centric approach is unknown.
 
To address this gap, in this paper we present an experience report on a code-centric approach to detect open source vulnerabilities using \codeAnalysis to determine whether vulnerable code is repackaged within Node.js applications.
First, we discuss the challenges associated with the construction of \codeAnalysis of Node.js applications.
We then propose our solution to counter these challenges.
To evaluate our approach, we perform a case study on 65 Node.js applications under development at SAP. 
Preliminary results show that our method is viable, with vulnerable code from five vulnerabilities being detected in 18 applications under development.
The study highlights three lessons learned and the challenges that require attention by both researchers and practitioners dealing with Node.js applications and JavaScript in general.
\begin{table*}[h]
\caption{Defined List of Constructs in hierarchical chain for Node.js Applications. This is based on Figure \ref{fig:project_structure} and Listing \ref{lst:util_b.js}}
\label{tab:constructTable}
\scalebox{0.9}{
\begin{tabular}{@{}lll@{}}
\toprule
\multicolumn{1}{c}{\textbf{Construct Type}} & \multicolumn{1}{c}{\textbf{Description}} & \multicolumn{1}{c}{\textbf{Fully Qualified Name}} \\ \midrule
Package (PACK) & Package and Directory name & ProjectA \\
Module (MODU) & File name & ProjectA.utils.util\_b \\
Function (FUNC) & Function name with arguments & ProjectA.utils.util\_b.buy(item) \\
Class (CLAS) & Class name with extended class & ProjectA.utils.util\_b.Car() \\
Method (METH) & Method name with arguments & ProjectA.utils.util\_b.Car().drive(distance,direction) \\
Constructor (CONS) & Constructor name with arguments & ProjectA.utils.util\_b.Car().constructor(name,age) \\
Object (OBJT) & Object name & ProjectA.utils.util\_b.item\_list \\ \bottomrule
\end{tabular}
}
\end{table*}

\section{Perils of JavaScript Node.js Analysis}
Analysis of JavaScript code is not trivial, as server-side Node.js applications (including npm packages) involve sockets, streams, and files performed in an asynchronous manner, where the execution of listeners is triggered by events \citep{Madsen:OOPSLA2015}.
The challenge for such dynamic code is the proper identification of a function call, which has been the issue for static analysis tools~\citep{Sung:FSE2016,Davis:EuroSys2017}.
Moreover, JavaScript allows anonymous functions, i.e., functions without a name~\citep{Web:AnonFunc}.
To avoid this complexity of the reachability analysis, our approach is based on the detection of vulnerable code.

We reuse the approach proposed by \citet{Ponta:icsme2018,Ponta:emse2020}, where a vulnerability
is detected whenever an application dependency contains \textit{program constructs} (such as methods) that were modified, added, or deleted to fix that vulnerability. 
We extend \tool~to support the analysis of JavaScript code \citep{NAISTSE:online}.
In particular, we add the ability to construct the list of program constructs modified to fix JavaScript vulnerabilities, as well as the list of program constructs which are part of a JavaScript application and dependencies (its bill of materials).

\section{\codeAnalysis for Node.js}
\label{sec:construct}
When compared to the classical model (i.e., Java or C++), JavaScript does not provide a true class implementation.
Instead, it has only the object construct with its private property (i.e., prototype) to imitate the constructs from the classical model \citep{Web:js_prototype_chain}.
A \texttt{program construct} is defined as a set of structural elements with a language, a type, and a unique fully-qualified name identifier as defined in~\citet{Ponta:icsme2018,Ponta:emse2020}. %

\subsection{Constructs for a Node.js application}

\begin{lstlisting}[language=JavaScript, caption={Example of a class of util\_b.js}, label={lst:util_b.js}]
class Car {
    constructor(name, age) { ...
    }
    drive(distance, direction) { ...
    }
}
var item_list = { ...
}
function buy(item) { ...
}
\end{lstlisting}

\begin{figure}[!h]
\centering
\includegraphics[width=0.4\textwidth]{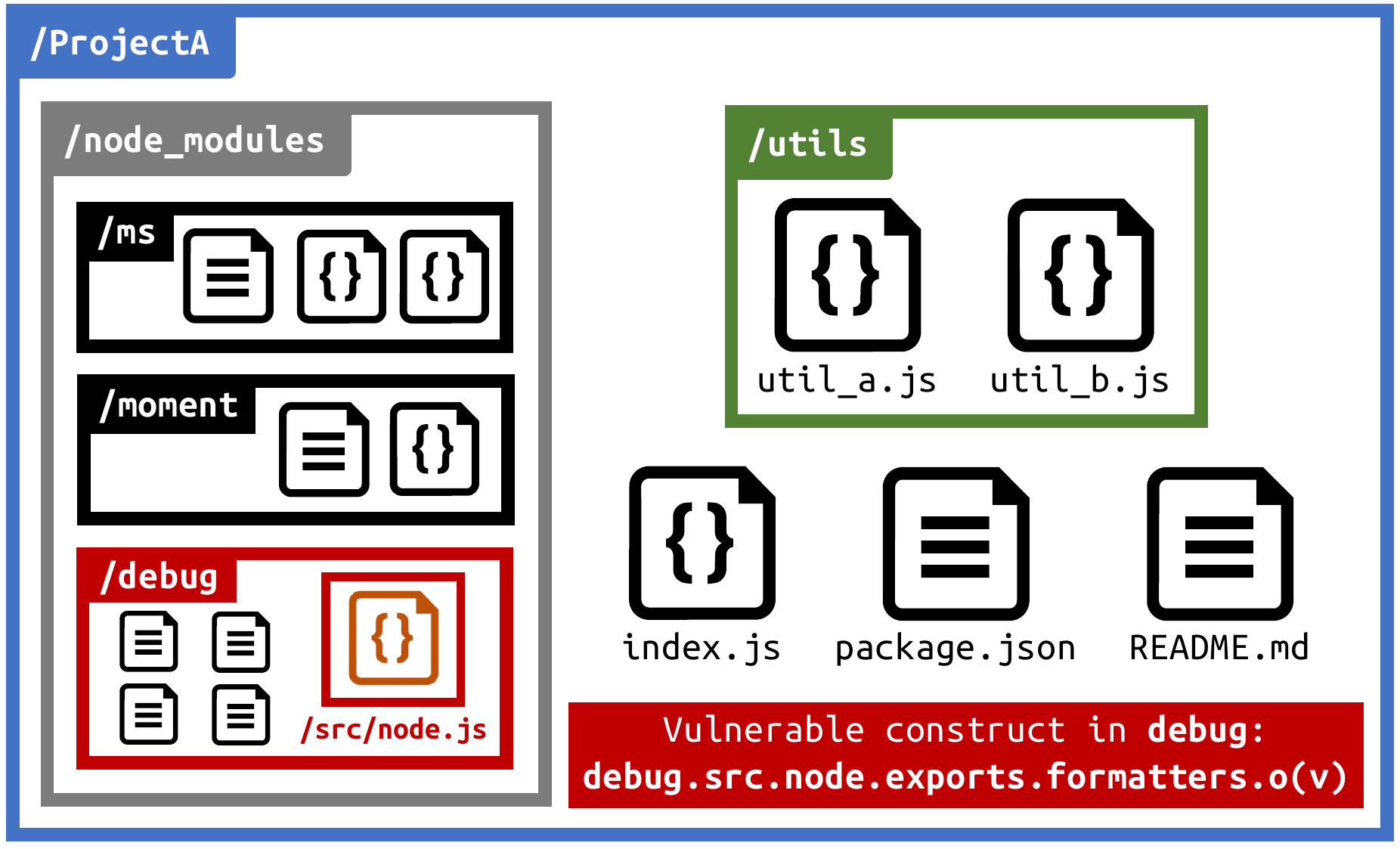}
\caption{Running example of the Node.js project with its hierarchical structure.}
\label{fig:project_structure}
\end{figure}

Figure \ref{fig:project_structure} illustrates a hierarchical structure of a Node.js project, which will be used as our running example.
Complementary, Listing \ref{lst:util_b.js} shows a code snippet from the JavaScript file \texttt{util\_b.js} of Figure~\ref{fig:project_structure}.
We use these running examples to explain our proposed constructs.

Table \ref{tab:constructTable} shows a summary of the seven construct types we use for Node.js applications.
We now explain each construct in detail.
The \texttt{PACK} construct represents an application scope and its internal directories e.g., \texttt{ProjectA}, \texttt{/utils}.
The \texttt{MODU} construct represents a JavaScript file in an application e.g., \texttt{util\_a.js}.
The \texttt{FUNC} construct represents a function declaration in a \texttt{MODU} e.g., \texttt{buy(item)}.
The \texttt{CLAS} construct represents a class declaration in a \texttt{MODU} e.g., \texttt{Car()}.
The \texttt{METH} construct represents a method declaration in a \texttt{CLAS} e.g., \texttt{drive(distance, direction)}.
The \texttt{CONS} construct represents a constructor declaration in a \texttt{CLAS} e.g., \texttt{constructor(name, age)}.
The \texttt{OBJT} construct represents an object in a \texttt{MODU} e.g., \texttt{item\_list}.

As shown in Table~\ref{tab:constructTable}, we use the PACK construct (i.e., ProjectA) hierarchy to form our fully qualified name.
Since JavaScript does allow for anonymous functions, classes, or objects, we use the (LoC) position for the fully-qualified name of anonymous constructs.

\subsection{Dependency Constructs and their Features}
The program constructs defined in Section~\ref{sec:construct} are also used to obtain the bill of materials of the third-party dependencies that are contained within the application (i.e., npm package).
Following our running example in Figure \ref{fig:project_structure}, we use the \texttt{package.json} configuration file and the \texttt{nodes\_modules} directory.
In our example, the vulnerable construct is in the debug package at the \texttt{OBJT} level (i.e., \sloppypar{\texttt{debug.src.node.exports.formatters.o(v)}})

\begin{lstlisting}[language=json, caption={Dependency snippet from package.json}, label={lst:package.json}]
 ...
  },
  "dependencies": {
    "moment": "2.25.3"
  },
  "devDependencies": {
    "debug": "3.0.0"
  },
  ...
\end{lstlisting}

Listing \ref{lst:package.json} shows that our example \texttt{ProjectA} depends on the packages \texttt{debug} and \texttt{moment}.
On top of collecting the bill of materials of each dependency, we analyze them according to two features.
The first is related to whether or not the dependency will be used in production.
There are two types of dependencies.
Runtime dependencies (i.e., \texttt{moment} package ) are those intended to be used in production. 
Test dependencies (i.e., \texttt{debug} package) are intended as development-only packages, unneeded in production. 

\begin{lstlisting}[language=bash, caption={Dependency tree of ProjectA}, label={lst:dependency}]
ProjectA@1.0.0 /ProjectA
|--- debug@4.1.1
| |--- ms@2.1.2
|--- moment@2.25.3
\end{lstlisting}

The second feature is the dependency tree depth.
Listing \ref{lst:dependency} shows the dependency tree that depicts the relationships among the packages \texttt{debug}, \texttt{moments} and \texttt{ms}.
There are two types of dependencies: direct and transitive. \textit{Direct} dependencies are directly required by the application. 
As shown in the example, the packages \texttt{debug} and \texttt{moment} are direct dependencies and are listed in the \texttt{package.json} file.
\textit{Transitive} dependencies are not directly required by the application but are required by its dependencies. 
As shown in Listing \ref{lst:dependency}, the \texttt{ms} package is a transitive package required by \texttt{debug}.

\begin{table}[b]
\caption{Experimental Dataset}
\label{tab:dataset}
\centering
\begin{tabular}{@{}lc@{}}
\toprule
\multicolumn{2}{c}{\textbf{OSS npm package vulnerabilities}} \\ \midrule
\multicolumn{1}{l|}{number of vulnerabilities} & 60 \\
\multicolumn{1}{l|}{number of vulnerable package} & 24 \\
\multicolumn{1}{l|}{number of valid vulnerabilities} & 32 \\
\multicolumn{1}{l|}{number of valid vulnerable package} & 15 \\ \midrule
\multicolumn{2}{c}{\textbf{Industrial Node.js applications}} \\ \midrule
\multicolumn{1}{l|}{number of applications} & 65 \\
\multicolumn{1}{l|}{number of valid application} & 42 \\ \bottomrule
\end{tabular}
\end{table}

\section{Case Study of Node.js Applications}
To evaluate our proposed constructs, we conducted an assessment of vulnerable code from under-development projects at \company.

\subsection{Experiment Design}
The experiment consisted of the detection of a set of known open source vulnerabilities against a set of industrial applications.
Note that our experiment was conducted in September, 2019.

\paragraph{\textbf{Bill of materials extraction}}
We create a bill of materials (BOM) which consists of construct lists of the application and its dependencies (i.e., both direct and transitive).
To do this, we use ANTLR-v4 with a JavaScript grammar \citep{Web:js_grammar}
to model and extract the Node.js application source code.
This grammar is able to partially extract JavaScript with ES6 features at the beginning of our development (July 24, 2019).

To obtain the BOM from an application, we first download the application dependencies by using \texttt{npm install}.
We then explore and build the dependency tree by looking at the \texttt{package-lock.json} file.
After that, we use ANTLR-v4 to extract the list of constructs from the JavaScript files of the application.
Next, we traverse the dependency tree depth-first to extract the list of constructs from each dependency.

\paragraph{\textbf{Vulnerability knowledge base}} 
We build our own Node.js vulnerability dataset which includes the vulnerability information and its fix for \tool.
We first retrieve the list of Node.js vulnerabilities and their information from the National Vulnerability Database (NVD) \citep{Web:NVD}.
We selected only vulnerabilities that have fixes and affect the top-100 most depended npm packages \citep{Web:npm_depended}.
We then manually annotate the set of commits that correspond to the vulnerability fix.
The set of commits has to be confirmed as it appeared on the master branch of the library git repository.
Given the fix commit(s) for a vulnerability, we use our extension of \tool~to determine the changes that were applied to the code by the fix commit.
As shown in Table \ref{tab:dataset}, we end up with 60 vulnerabilities in our study.

\paragraph{\textbf{SAP Node.js Applications}}
We used \company~GitHub enterprise to identify Node.js applications suitable for our case study. We considered only applications under development having the \texttt{package.json} file in their root directory. We selected a sample of 65 applications, as shown in Table \ref{tab:dataset}.

\begin{table}[h]
\centering
\caption{Dependency Type information.}
\label{tab:dependency_type_info}
\scalebox{0.95}{
\begin{tabular}{@{}lrrrrrr@{}}
\toprule
\multicolumn{1}{c}{\textbf{\# Dependencies}} & \multicolumn{1}{c}{\textbf{Median}} & \multicolumn{1}{c}{\textbf{Min}} & \multicolumn{1}{c}{\textbf{Max}} & \multicolumn{1}{c}{\textbf{Q1}} & \multicolumn{1}{c}{\textbf{Q3}} & \multicolumn{1}{c}{\textbf{SD}} \\ \midrule
\# All Dep. & 464.5 & 3 & 1,226 & 229.75 & 748.5 & 339.55 \\
\# Runtime Dep. & 108.5 & 0 & 586 & 40.75 & 193 & 146.18 \\
\# Test Dep. & 257 & 0 & 1,067 & 117.25 & 561.5 & 335.31 \\ \bottomrule
\end{tabular}
}
\end{table}

Table \ref{tab:dependency_type_info} shows the distribution of dependencies, showing more than a hundred dependencies in each application by median (i.e. 464.5 dependencies) with some applications having up to a thousand dependencies (i.e., 1,226 dependencies).
We observe that the number of test dependencies is bigger than the one of runtime dependencies by two times (i.e., 257 > 108.5).

\subsection{Results}
We present our results in terms of: (i) detected vulnerabilities, and (ii) dependency constructs.

\paragraph{\textbf{Detected Vulnerabilities}}
Our prototype was able to detect five vulnerabilities that affected the \texttt{lodash} and \texttt{debug} npm packages.
Lodash \citep{Web:lodash}
is "A modern JavaScript utility library delivering modularity, performance, and extras".
According to the npmjs website \citep{Web:npm_lodash},
lodash is a very popular package, with over 27,500,000 weekly downloads and 114,917 other packages that are dependent on this package.
Debug \citep{Web:debug}
is "A tiny JavaScript debugging utility modelled after Node.js core's debugging technique".
According to the npmjs website \citep{Web:npm_debug},
debug is also considered a popular package, with over 66,800,000 weekly downloads and 34,494 dependents.

\paragraph{\textbf{Dependency Constructs and features}}

In our case study, our extension to \tool could analyze 42 out of 65 applications.

Table \ref{tab:construct_type_info} shows the distribution of the BOM extracted from the applications.
Our prototype was able to extract more than a hundred constructs from an application and its dependencies (i.e., 164.5 constructs).
In more detail, the number of application constructs is bigger than the one of dependency constructs by three times (i.e., 75 > 26).

\begin{table}[h]
\centering
\caption{Summary of Construct Information from the experiment.}
\label{tab:construct_type_info}
\scalebox{0.9}{
\begin{tabular}{@{}lrrrrrr@{}}
\toprule
\multicolumn{1}{c}{\textbf{\# Constructs}} & \multicolumn{1}{c}{\textbf{Median}} & \multicolumn{1}{c}{\textbf{Min}} & \multicolumn{1}{c}{\textbf{Max}} & \multicolumn{1}{c}{\textbf{Q1}} & \multicolumn{1}{c}{\textbf{Q3}} & \multicolumn{1}{c}{\textbf{SD}} \\ \midrule
\# App Consts. & 75 & 0 & 3,083 & 28.25 & 167.5 & 573.99 \\
\# Dep Consts. & 26 & 0 & 9,549 & 1.25 & 114.25 & 2,144.69 \\
\# App + Dep Consts. & 164.5 & 1 & 9,671 & 83.25 & 609.75 & 2,224.14 \\ \bottomrule
\end{tabular}
}
\vspace{-1em}
\end{table}

\begin{table}[h]
\centering
\caption{Frequency count of Dependent Construct Changes per vulnerability}
\label{tab:construct_change_distribution}
\scalebox{0.9}{
\begin{tabular}{@{}llll@{}}
\toprule
\multicolumn{1}{c}{\multirow{2}{*}{\textbf{CVE}}} & \multicolumn{3}{c}{\textbf{Construct Change Type}} \\ \cmidrule(l){2-4} 
\multicolumn{1}{c}{} & \multicolumn{1}{c}{\textbf{Added}} & \multicolumn{1}{c}{\textbf{Modified}} & \multicolumn{1}{c}{\textbf{Removed}} \\ \midrule
CVE-2017-16137 & FUNC:1 & MODU:1, FUNC:1 &  \\
CVE-2018-3721 & FUNC:2, OBJT:1 & MODU:2, FUNC:7 &  \\
CVE-2018-16487 & FUNC:2, OBJT:4 & MODU:2, FUNC:4 & FUNC:1 \\
CVE-2019-10744 & FUNC:1, OBJT:3 & MODU:2, FUNC:5 &  \\
CVE-2019-1010266 & FUNC:1 & MODU:2, FUNC:3 &  \\ \bottomrule
\end{tabular}
}
\vspace{-1em}
\end{table}

\begin{table}[h]
\caption{Frequency distribution of Dependency Constructs based on the dependency features.}
\label{tab:affected_application_distribution}
\centering
\scalebox{0.9}{
\begin{tabular}{@{}lrrrr@{}}
\toprule
\multicolumn{1}{c}{\multirow{2}{*}{\textbf{Vulnerability}}} & \multicolumn{2}{c}{\textbf{Runtime (26)}} & \multicolumn{2}{c}{\textbf{Test (31)}} \\ \cmidrule(l){2-5} 
\multicolumn{1}{c}{} & \multicolumn{1}{c}{\textbf{Direct}} & \multicolumn{1}{c}{\textbf{Trans.}} & \multicolumn{1}{c}{\textbf{Direct}} & \multicolumn{1}{c}{\textbf{Trans.}} \\ \midrule
CVE-2017-16137 & 0 & 1 & 0 & 11 \\
CVE-2018-3721 & 0 & 6 & 1 & 2 \\
CVE-2018-16487 & 0 & 6 & 1 & 3 \\
CVE-2019-10744 & 0 & 7 & 1 & 8 \\
CVE-2019-1010266 & 0 & 6 & 1 & 3 \\ \bottomrule
& 0 & 26 & 4 & 27 \\ \bottomrule
\end{tabular}
}
\vspace{-1em}
\end{table}

Table \ref{tab:construct_change_distribution} and Table \ref{tab:affected_application_distribution}  show the affected dependency constructs and their construct type.
Table \ref{tab:construct_change_distribution} shows that the construct changes were detected at the \texttt{OBJT}, \texttt{MODU} and \texttt{FUNC} level.   
We observe that the majority of the vulnerable dependencies are transitive (28 runtime dependencies and 27 test dependencies), i.e., usually out of the control of the application developer.
We also observe that most of the vulnerable constructs are detected in test dependencies.

\section{Experience Report}
Our results indicate that a Node.js vulnerable code detector is viable.
We now report three lessons learned and their potential future roadmap.

\subsection{Mapping JavaScript Object to Constructs}
In our approach, we defined a more classical inheritance of constructs (like Java and C++) on top of the JavaScript prototypal inheritance model.
With this choice, one of the main issues is to ensure that we capture all the different ways to create objects and their constructs.
For instance, there are at least six way to declare a function in JavaScript \citep{Web:js_funciton_declaration}.
Furthermore, it is still an open question whether the implementation efforts required to extract the finer-level constructs (e.g., OBJT) are worth. %
As shown in Table \ref{tab:construct_change_distribution}, in most of the cases the \texttt{MODU} constructs were sufficient for the detection of the vulnerabilities.

Potential future avenues are two-fold. 
First, we would like to consider all the ways in which objects can be created in JavaScript.
Second, we intend to evaluate the detection capabilities of our approach at different levels of the construct hierarchy (i.e., MODU vs. FUNC vs. OBJT).

\subsection{\textbf{Node.js application reliance on the npm ecosystem}}
The applications in our case study rely on npm packages, and as such, are potentially prone to attacks targeting popular packages, like the \texttt{lodash} and \texttt{debug} packages.
Since the npm ecosystem is considered one of the biggest and most popular, it does also suffer the most in terms of known vulnerabilities, with the GitHub Advisory Database reporting npm as having the highest number of vulnerabilities (i.e., 681) when compared to six other ecosystems~\citep{Web:github_advisory}.

With the GitHub acquisition of npm, we envisage that Node.js applications will need to be aware of changes within the npm ecosystem.
The creation and evaluation of such reporting mechanisms are seen a future work.

\subsection{Faster Technology Adoption}
Officially known as ECMAScript, the JavaScript language has been in constant evolution with its technology, with new specifications released every year.
In response, Node.js keeps up to date~\citep{Web:NodejsES}.
Since industrial projects struggle with migration due to various migration or compatibility issue, it is a struggle for applications to keep up with the Node.js technology.
For example, practitioners would like to control or specify the supported platform of the language.
As mentioned in Section 5.2, the usage of npm packages requires industrial applications to keep up with the npm ecosystem evolution.
Like most tools, we find that JavaScript static tools (such as ANTLRv4) struggle to keep up to date.

Potential future avenues for both researchers and practitioners should include strategies that help application developers to properly manage backward compatibility or guidelines to keep up with the ever-evolving technology.

\section{Conclusion}
\label{sec:conclusion}
In this paper, we present an experience report on the implementation of a code-centric vulnerability detection tool for open source dependencies of Node.js applications.
Using extracted constructs, we show that a code-centric detection tool is viable, although there are challenges related to the JavaScript language and the complexity of the application dependencies.

Future work would be to tackle the challenges of JavaScript analysis, or extending the tool to analyze the reachability of vulnerable constructs using static and dynamic analysis techniques.
We believe that our results and experience is not only useful for the \tool~project, but also in regards to the overall analysis of Node.js applications and their npm packages.

\section*{Acknowledgement}
This work is supported by Japanese Society for the Promotion of Science (JSPS) KAKENHI Grant Numbers 18H04094, 18H03221, 18KT0013, and 20K19774.

\bibliographystyle{ACM-Reference-Format}
\bibliography{references}


\begin{thebibliography}{29}


\ifx \showCODEN    \undefined \def \showCODEN     #1{\unskip}     \fi
\ifx \showDOI      \undefined \def \showDOI       #1{#1}\fi
\ifx \showISBNx    \undefined \def \showISBNx     #1{\unskip}     \fi
\ifx \showISBNxiii \undefined \def \showISBNxiii  #1{\unskip}     \fi
\ifx \showISSN     \undefined \def \showISSN      #1{\unskip}     \fi
\ifx \showLCCN     \undefined \def \showLCCN      #1{\unskip}     \fi
\ifx \shownote     \undefined \def \shownote      #1{#1}          \fi
\ifx \showarticletitle \undefined \def \showarticletitle #1{#1}   \fi
\ifx \showURL      \undefined \def \showURL       {\relax}        \fi
\providecommand\bibfield[2]{#2}
\providecommand\bibinfo[2]{#2}
\providecommand\natexlab[1]{#1}
\providecommand\showeprint[2][]{arXiv:#2}

\bibitem[\protect\citeauthoryear{ANTLR}{ANTLR}{2017}]%
        {Web:js_grammar}
\bibfield{author}{\bibinfo{person}{ANTLR}.} \bibinfo{year}{2017}\natexlab{}.
\newblock \bibinfo{title}{grammars-v4/javascript at master ·
  antlr/grammars-v4}.
\newblock
  \bibinfo{howpublished}{\url{https://github.com/antlr/grammars-v4/tree/master/javascript}}.
\newblock
\newblock
\shownote{(Accessed on 08/11/2020).}


\bibitem[\protect\citeauthoryear{Database}{Database}{2007}]%
        {Web:NVD}
\bibfield{author}{\bibinfo{person}{National~Vulnerability Database}.}
  \bibinfo{year}{2007}\natexlab{}.
\newblock \bibinfo{title}{NVD - Home}.
\newblock \bibinfo{howpublished}{\url{https://nvd.nist.gov/}}.
\newblock
\newblock
\shownote{(Accessed on 08/11/2020).}


\bibitem[\protect\citeauthoryear{Davis, Thekumparampil, and Lee}{Davis
  et~al\mbox{.}}{2017}]%
        {Davis:EuroSys2017}
\bibfield{author}{\bibinfo{person}{James Davis}, \bibinfo{person}{Arun
  Thekumparampil}, {and} \bibinfo{person}{Dongyoon Lee}.}
  \bibinfo{year}{2017}\natexlab{}.
\newblock \showarticletitle{Node.Fz: Fuzzing the Server-Side Event-Driven
  Architecture}. In \bibinfo{booktitle}{\emph{Proceedings of the 12th European
  Conference on Computer Systems (EuroSys)}}. \bibinfo{pages}{145–160}.
\newblock


\bibitem[\protect\citeauthoryear{{Decan}, {Mens}, and {Constantinou}}{{Decan}
  et~al\mbox{.}}{2018}]%
        {Decan:ICSME:2018}
\bibfield{author}{\bibinfo{person}{Alexandre {Decan}}, \bibinfo{person}{Tom
  {Mens}}, {and} \bibinfo{person}{Eleni {Constantinou}}.}
  \bibinfo{year}{2018}\natexlab{}.
\newblock \showarticletitle{On the Evolution of Technical Lag in the npm
  Package Dependency Network}. In \bibinfo{booktitle}{\emph{the 34th
  International Conference on Software Maintenance and Evolution (ICSME)}}.
  \bibinfo{pages}{404--414}.
\newblock


\bibitem[\protect\citeauthoryear{Decan, Mens, and Constantinou}{Decan
  et~al\mbox{.}}{2018}]%
        {Decan:2018}
\bibfield{author}{\bibinfo{person}{Alexandre Decan}, \bibinfo{person}{Tom
  Mens}, {and} \bibinfo{person}{Eleni Constantinou}.}
  \bibinfo{year}{2018}\natexlab{}.
\newblock \showarticletitle{{On the impact of security vulnerabilities in the
  npm package dependency network}}. In \bibinfo{booktitle}{\emph{Proceedings of
  the 15th International Conference on Mining Software Repositories (MSR)}}.
  \bibinfo{pages}{181--191}.
\newblock


\bibitem[\protect\citeauthoryear{Eclipse}{Eclipse}{2018}]%
        {Web:Steady}
\bibfield{author}{\bibinfo{person}{Eclipse}.} \bibinfo{year}{2018}\natexlab{}.
\newblock \bibinfo{title}{Eclipse Steady 3.1.11 (Incubator Project)}.
\newblock \bibinfo{howpublished}{\url{https://eclipse.github.io/steady/}}.
\newblock
\newblock
\shownote{(Accessed on 08/11/2020).}


\bibitem[\protect\citeauthoryear{GitHub}{GitHub}{2017}]%
        {Web:github_security_noti}
\bibfield{author}{\bibinfo{person}{GitHub}.} \bibinfo{year}{2017}\natexlab{}.
\newblock \bibinfo{title}{{About security alerts for vulnerable dependencies}}.
\newblock
  \bibinfo{howpublished}{\url{https://help.github.com/articles/about-security-alerts-for-vulnerable-dependencies/}}.
\newblock
\newblock
\shownote{(Accessed on 08/11/2020).}


\bibitem[\protect\citeauthoryear{GitHub}{GitHub}{2019}]%
        {Web:github_advisory}
\bibfield{author}{\bibinfo{person}{GitHub}.} \bibinfo{year}{2019}\natexlab{}.
\newblock \bibinfo{title}{GitHub Advisory Database}.
\newblock \bibinfo{howpublished}{\url{https://github.com/advisories}}.
\newblock
\newblock
\shownote{(Accessed on 08/11/2020).}


\bibitem[\protect\citeauthoryear{Kapke}{Kapke}{2016}]%
        {Web:NodejsES}
\bibfield{author}{\bibinfo{person}{William Kapke}.}
  \bibinfo{year}{2016}\natexlab{}.
\newblock \bibinfo{title}{Node.js ES2015/ES6, ES2016 and ES2017 support}.
\newblock \bibinfo{howpublished}{\url{https://node.green/}}.
\newblock
\newblock
\shownote{(Accessed on 08/11/2020).}


\bibitem[\protect\citeauthoryear{Kikas, Gousios, Dumas, and Pfahl}{Kikas
  et~al\mbox{.}}{2017}]%
        {Kikas:2017}
\bibfield{author}{\bibinfo{person}{Riivo Kikas}, \bibinfo{person}{Georgios
  Gousios}, \bibinfo{person}{Marlon Dumas}, {and} \bibinfo{person}{Dietmar
  Pfahl}.} \bibinfo{year}{2017}\natexlab{}.
\newblock \showarticletitle{{Structure and Evolution of Package Dependency
  Networks}}. In \bibinfo{booktitle}{\emph{Proceedings of the 14th
  International Conference on Mining Software Repositories (MSR)}}.
  \bibinfo{pages}{102--112}.
\newblock


\bibitem[\protect\citeauthoryear{Lauinger, Chaabane, Arshad, Robertson, Wilson,
  and Kirda}{Lauinger et~al\mbox{.}}{2017}]%
        {Lauinger2016ThouSN}
\bibfield{author}{\bibinfo{person}{Tobias Lauinger}, \bibinfo{person}{Abdelberi
  Chaabane}, \bibinfo{person}{Sajjad Arshad}, \bibinfo{person}{William
  Robertson}, \bibinfo{person}{Christo Wilson}, {and} \bibinfo{person}{Engin
  Kirda}.} \bibinfo{year}{2017}\natexlab{}.
\newblock \showarticletitle{{Thou Shalt Not Depend on Me: Analysing the Use of
  Outdated JavaScript Libraries on the Web}}. In
  \bibinfo{booktitle}{\emph{Proceedings of the 24th Network and Distributed
  System Security Symposium (NDSS)}}.
\newblock


\bibitem[\protect\citeauthoryear{Lodash}{Lodash}{2012}]%
        {Web:lodash}
\bibfield{author}{\bibinfo{person}{Lodash}.} \bibinfo{year}{2012}\natexlab{}.
\newblock \bibinfo{title}{lodash/lodash: A modern JavaScript utility library
  delivering modularity, performance, \& extras.}
\newblock \bibinfo{howpublished}{\url{https://github.com/lodash/lodash}}.
\newblock
\newblock
\shownote{(Accessed on 08/11/2020).}


\bibitem[\protect\citeauthoryear{Madsen, Tip, and Lhot\'{a}k}{Madsen
  et~al\mbox{.}}{2015}]%
        {Madsen:OOPSLA2015}
\bibfield{author}{\bibinfo{person}{Magnus Madsen}, \bibinfo{person}{Frank Tip},
  {and} \bibinfo{person}{Ond\v{r}ej Lhot\'{a}k}.}
  \bibinfo{year}{2015}\natexlab{}.
\newblock \showarticletitle{Static Analysis of Event-Driven Node.Js JavaScript
  Applications}. In \bibinfo{booktitle}{\emph{Proceedings of the International
  Conference on Object-Oriented Programming, Systems, Languages, and
  Applications (OOPSLA)}}. \bibinfo{pages}{505–519}.
\newblock


\bibitem[\protect\citeauthoryear{NAIST-SE}{NAIST-SE}{2020}]%
        {NAISTSE:online}
\bibfield{author}{\bibinfo{person}{NAIST-SE}.} \bibinfo{year}{2020}\natexlab{}.
\newblock \bibinfo{title}{NAIST-SE/steady: Analyses your Java and Python
  applications for open-source dependencies with known vulnerabilities, using
  both static analysis and testing to determine code context and usage for
  greater accuracy.}
\newblock \bibinfo{howpublished}{\url{https://github.com/NAIST-SE/steady}}.
\newblock
\newblock
\shownote{(Accessed on 08/11/2020).}


\bibitem[\protect\citeauthoryear{npm}{npm}{2011}]%
        {Web:npm_debug}
\bibfield{author}{\bibinfo{person}{npm}.} \bibinfo{year}{2011}\natexlab{}.
\newblock \bibinfo{title}{debug - npm}.
\newblock \bibinfo{howpublished}{\url{https://www.npmjs.com/package/debug}}.
\newblock
\newblock
\shownote{(Accessed on 08/11/2020).}


\bibitem[\protect\citeauthoryear{npm}{npm}{2012}]%
        {Web:npm_lodash}
\bibfield{author}{\bibinfo{person}{npm}.} \bibinfo{year}{2012}\natexlab{}.
\newblock \bibinfo{title}{lodash - npm}.
\newblock \bibinfo{howpublished}{\url{https://www.npmjs.com/package/lodash}}.
\newblock
\newblock
\shownote{(Accessed on 08/11/2020).}


\bibitem[\protect\citeauthoryear{{NPM}}{{NPM}}{2018}]%
        {Web:npm_audit}
\bibfield{author}{\bibinfo{person}{{NPM}}.} \bibinfo{year}{2018}\natexlab{}.
\newblock \bibinfo{title}{{Auditing package dependencies for security
  vulnerabilities}}.
\newblock
  \bibinfo{howpublished}{\url{https://docs.npmjs.com/auditing-package-dependencies-for-security-vulnerabilities}}.
\newblock
\newblock
\shownote{(Accessed on 08/11/2020).}


\bibitem[\protect\citeauthoryear{npm}{npm}{2020}]%
        {Web:npm_depended}
\bibfield{author}{\bibinfo{person}{npm}.} \bibinfo{year}{2020}\natexlab{}.
\newblock \bibinfo{title}{npm - most dependend upon}.
\newblock \bibinfo{howpublished}{\url{https://www.npmjs.com/browse/depended}}.
\newblock
\newblock
\shownote{(Accessed on 08/11/2020).}


\bibitem[\protect\citeauthoryear{npm blog}{npm blog}{2020}]%
        {Web:npmStat}
\bibfield{author}{\bibinfo{person}{npm blog}.} \bibinfo{year}{2020}\natexlab{}.
\newblock \bibinfo{title}{{npm blog: Next Phase Montage}}.
\newblock
  \bibinfo{howpublished}{\url{https://blog.npmjs.org/post/612764866888007680/next-phase-montage}}.
\newblock
\newblock
\shownote{(Accessed on 05/20/2020).}


\bibitem[\protect\citeauthoryear{Pavlutin}{Pavlutin}{2016}]%
        {Web:js_funciton_declaration}
\bibfield{author}{\bibinfo{person}{Dmitri Pavlutin}.}
  \bibinfo{year}{2016}\natexlab{}.
\newblock \bibinfo{title}{6 Ways to Declare JavaScript Functions}.
\newblock
  \bibinfo{howpublished}{\url{https://dmitripavlutin.com/6-ways-to-declare-javascript-functions/}}.
\newblock
\newblock
\shownote{(Accessed on 08/11/2020).}


\bibitem[\protect\citeauthoryear{Ponta, Plate, and Sabetta}{Ponta
  et~al\mbox{.}}{2018}]%
        {Ponta:icsme2018}
\bibfield{author}{\bibinfo{person}{Serena~Elisa Ponta}, \bibinfo{person}{Henrik
  Plate}, {and} \bibinfo{person}{Antonino Sabetta}.}
  \bibinfo{year}{2018}\natexlab{}.
\newblock \showarticletitle{{Beyond Metadata: Code-centric and Usage-based
  Analysis of Known Vulnerabilities in Open-source Software}}. In
  \bibinfo{booktitle}{\emph{Proceedings of the 34th International Conference on
  Software Maintenance and Evolution (ICSME)}}. \bibinfo{pages}{58--68}.
\newblock


\bibitem[\protect\citeauthoryear{Ponta, Plate, and Sabetta}{Ponta
  et~al\mbox{.}}{2020}]%
        {Ponta:emse2020}
\bibfield{author}{\bibinfo{person}{Serena~Elisa Ponta}, \bibinfo{person}{Henrik
  Plate}, {and} \bibinfo{person}{Antonino Sabetta}.}
  \bibinfo{year}{2020}\natexlab{}.
\newblock \showarticletitle{{ Detection, assessment and mitigation of
  vulnerabilities in open source dependencies.}}
\newblock \bibinfo{journal}{\emph{Empirical Software Engineering (EMSE)}}
  (\bibinfo{year}{2020}).
\newblock
\urldef\tempurl%
\url{https://doi.org/10.1007/s10664-020-09830-x}
\showDOI{\tempurl}


\bibitem[\protect\citeauthoryear{Sung, Kusano, Sinha, and Wang}{Sung
  et~al\mbox{.}}{2016}]%
        {Sung:FSE2016}
\bibfield{author}{\bibinfo{person}{Chungha Sung}, \bibinfo{person}{Markus
  Kusano}, \bibinfo{person}{Nishant Sinha}, {and} \bibinfo{person}{Chao Wang}.}
  \bibinfo{year}{2016}\natexlab{}.
\newblock \showarticletitle{Static DOM Event Dependency Analysis for Testing
  Web Applications}. In \bibinfo{booktitle}{\emph{Proceedings of the 24th ACM
  SIGSOFT International Symposium on Foundations of Software Engineering
  (FSE)}}. \bibinfo{pages}{447–459}.
\newblock


\bibitem[\protect\citeauthoryear{Synopsys}{Synopsys}{2020}]%
        {Web:ossra}
\bibfield{author}{\bibinfo{person}{Synopsys}.} \bibinfo{year}{2020}\natexlab{}.
\newblock \bibinfo{title}{2020 Open Source Security and Risk Analysis (OSSRA)
  Report | Synopsys}.
\newblock
  \bibinfo{howpublished}{\url{https://www.synopsys.com/software-integrity/resources/analyst-reports/2020-open-source-security-risk-analysis.html}}.
\newblock
\newblock
\shownote{(Accessed on 05/27/2020).}


\bibitem[\protect\citeauthoryear{Tutorial}{Tutorial}{2020}]%
        {Web:AnonFunc}
\bibfield{author}{\bibinfo{person}{JavaScript Tutorial}.}
  \bibinfo{year}{2020}\natexlab{}.
\newblock \bibinfo{title}{JavaScript Anonymous Functions}.
\newblock
  \bibinfo{howpublished}{\url{https://www.javascripttutorial.net/javascript-anonymous-functions/}}.
\newblock
\newblock
\shownote{(Accessed on 08/11/2020).}


\bibitem[\protect\citeauthoryear{Visionmedia}{Visionmedia}{2011}]%
        {Web:debug}
\bibfield{author}{\bibinfo{person}{Visionmedia}.}
  \bibinfo{year}{2011}\natexlab{}.
\newblock \bibinfo{title}{visionmedia/debug: A tiny JavaScript debugging
  utility modelled after Node.js core's debugging technique. Works in Node.js
  and web browsers}.
\newblock \bibinfo{howpublished}{\url{https://github.com/visionmedia/debug}}.
\newblock
\newblock
\shownote{(Accessed on 08/11/2020).}


\bibitem[\protect\citeauthoryear{web docs}{web docs}{2020}]%
        {Web:js_prototype_chain}
\bibfield{author}{\bibinfo{person}{MDN web docs}.}
  \bibinfo{year}{2020}\natexlab{}.
\newblock \bibinfo{title}{Inheritance and the prototype chain - JavaScript |
  MDN}.
\newblock
  \bibinfo{howpublished}{\url{https://developer.mozilla.org/en-US/docs/Web/JavaScript/Inheritance_and_the_prototype_chain}}.
\newblock
\newblock
\shownote{(Accessed on 08/11/2020).}


\bibitem[\protect\citeauthoryear{Zapata, Kula, Chinthanet, Ishio, Matsumoto,
  and Ihara}{Zapata et~al\mbox{.}}{2018}]%
        {Zapata:ICSME2018}
\bibfield{author}{\bibinfo{person}{Rodrigo~Elizalde Zapata},
  \bibinfo{person}{Raula~Gaikovina Kula}, \bibinfo{person}{Bodin Chinthanet},
  \bibinfo{person}{Takashi Ishio}, \bibinfo{person}{Kenichi Matsumoto}, {and}
  \bibinfo{person}{Akinori Ihara}.} \bibinfo{year}{2018}\natexlab{}.
\newblock \showarticletitle{Towards Smoother Library Migrations: A Look at
  Vulnerable Dependency Migrations at Function Level for npm JavaScript
  Packages}. In \bibinfo{booktitle}{\emph{Proceedings of the 34th International
  Conference on Software Maintenance and Evolution (ICSME)}}.
  \bibinfo{pages}{559--563}.
\newblock


\bibitem[\protect\citeauthoryear{Zerouali, Constantinou, Mens, Robles, and
  Gonzalez-Barahona}{Zerouali et~al\mbox{.}}{2018}]%
        {Zerouali:ICSR2018}
\bibfield{author}{\bibinfo{person}{Ahmed Zerouali}, \bibinfo{person}{Eleni
  Constantinou}, \bibinfo{person}{Tom Mens}, \bibinfo{person}{Gregorio Robles},
  {and} \bibinfo{person}{Jesus Gonzalez-Barahona}.}
  \bibinfo{year}{2018}\natexlab{}.
\newblock \showarticletitle{An Empirical Analysis of Technical Lag in npm
  Package Dependencies}. In \bibinfo{booktitle}{\emph{Proceedings of the 17th
  International Conference on Software Reuse (ICSR)}}.
  \bibinfo{pages}{95--110}.
\newblock


\end{thebibliography}

\end{document}